\documentclass[aps,prev,onecolumn,preprintnumbers,floatfix,nofootinbib,11pt]{revtex4}
\def\lsim{\raise0.3ex\hbox{$\;<$\kern-0.75em\raise-1.1ex\hbox{$\sim\;$}}}
\def\gsim{\raise0.3ex\hbox{$\;>$\kern-0.75em\raise-1.1ex\hbox{$\sim\;$}}}

\def\ie{{\it i.e.}}
\usepackage{graphicx}
\usepackage{multirow}
\usepackage{pstricks}
\usepackage{dcolumn}
\usepackage{epsfig}
\usepackage{amsmath}
\usepackage{amsfonts}
\usepackage{float}
\usepackage{amssymb}
\usepackage{color}

\newcommand{\mat}[1]{\begin{pmatrix} #1 \end{pmatrix}}

\usepackage{graphics}

\usepackage{epsfig}
\usepackage{slashed}
\usepackage[utf8]{inputenc}
\newcommand{\be}{\begin{eqnarray}}
\newcommand{\ee}{\end{eqnarray}}

\def\bea{\begin{eqnarray}}
\def\eea{\end{eqnarray}}
\usepackage{epsfig}
\usepackage{array}
\usepackage{slashed}

\begin{document}
\title{Explaining the $R_{K}$ and $R_{K^*}$ Anomalies With Right-handed Sneutrino}
\author{Shaaban Khalil}
\vspace*{0.2cm}
\affiliation{Center for Fundamental Physics, Zewail City of Science and Technology, 6 October City,
Giza 12588, Egypt.}
\date{\today}

\begin{abstract}
The recent intriguing measurements of $R_{K}$ and $R_{K^*}$ are important hints of new physics that violates lepton
universality. We analyze the semileptonic decays $b \to s \ell^+ \ell^-$ in the framework of the $B-L$ extension of Minimal Supersymmetric
Standard Model (MSSM) with Inverse Seesaw (BLSSMIS). A salient feature of this model  is that one of the right-handed sneutrino can be light and the neutrino Yukawa couplings are of order one. We show that the box diagram mediated by right-handed sneutrino, higgsino-like chargino, and light stop can account simultaneously for both $R_{K}$ and $R_{K^*}$.
Therefore, while the MSSM cannot explain the $R_K$ and $R_{K^*}$ anomalies, the BLSSMIS can account for them, where a significant lepton flavor non-universality might stem from large neutrino Yukawa couplings.

\end{abstract}
\maketitle

%%%%%%%%%%%%%%%%%%%%%%%%%%%%%%%
\section{Introduction}
Flavor Changing Neutral Currents (FCNCs) are considered as the best indirect probes for physics beyond the Standard Model (SM).
They are particularly sensitive to New Physics (NP), due to their very large suppressions in the SM. Recently, the LHCb collaboration has reported an interesting result \cite{Aaij:2014ora} for the ratio $R_{K}= {\rm BR}(B^+\to K^+ \mu^+\mu^-)/{\rm BR}(B^+\to K^+ e^+ e^-)$. They found that for dilepton invariant mass-squared $1 \leq q^2 \leq 6$ GeV$^2$, $R_{K}$ is given by
\be
R_{K} = 0.745~ ^{+0.090}_{-0.074} ~\pm ~ 0.036.
\label{RKexp}
\ee
This measurement of lepton non-universality parameter differs from the SM expectation: $R_{K}^{\rm SM}= 1\pm 0.01$ \cite{Bordone:2016gaq} by $2.6 \sigma$. This result is similar to another finding by LHCb collaboration \cite{Aaij:2017vbb}, where they measured the ratio $R_{K^*}= {\rm BR}(B^0\to K^{*0} \mu^+\mu^-)/{\rm BR}(B^0\to K^{*0} e^+ e^-)$ and found that
\be
R_{K^*} = 0.69~ ^{+0.11}_{-0.07} ~\pm ~ 0.05 ~~~~~~ ~~~~~~ 1.1~ {\rm GeV} ^2 < q^2 < 6.0 ~{\rm GeV}^2,
\label{RK*exp}
\ee
which again is less than the SM prediction: $R_{K^*}^{\rm SM}\simeq 1$ \cite{Bordone:2016gaq} by $2.5 \sigma$.
The theoretical uncertainties in calculating ${\rm BR}(B\to K \ell^+ \ell^-)$ is essentially canceled in both $R_{K}$ and
$R_{K^*}$. Thus, confirming these discrepancies would make them very clean signal of NP~\cite{Hiller:2003js}. Recently, there has been growing interest in exploring  NP scenarios that may explain these anomalies \cite{Ghosh:2017ber,Capdevila:2017bsm,Altmannshofer:2017yso,DAmico:2017mtc,Hiller:2017bzc,Geng:2017svp,Ciuchini:2017mik,Celis:2017doq,Becirevic:2017jtw,Alok:2017sui,Feruglio:2017rjo,Alok:2017jaf,Wang:2017mrd,Ellis:2017nrp,Chiang:2017hlj,King:2017anf,Datta:2017ezo,Crivellin:2015lwa,Arnan:2016cpy,Crivellin:2017zlb,Bonilla:2017lsq,Chauhan:2017ndd,Das:2017kfo,Bardhan:2017xcc}. In this paper we argue that the BLSSMIS one-loop box diagram, generated by right-handed sneutrino, light stop, and higgsino-like chargino, 
can account for the discrepancy between the experimental results of $R_K \,\&\, R_{K^*}$ and  SM expectations. 
It is worth mentioning that in the MSSM  large contributions to the non-universal lepton processes are not allowed \cite{Mahmoudi:2014mja,Altmannshofer:2014rta}. Even in the MSSM with $R$-parity violating interactions, it was explicitly shown that enhancing the corresponding loop contribution to $b \to s$ anomalies is restricted by the experimental constrained on the $R$-parity violating coupling and also the mass of the sfermions \cite{Earl:2018snx}. Therefore, one concludes the MSSM predictions for both $R_K$ and $R_{K^*}$ are almost identical to their SM values, hence it can not account for the recent observed anomalies. To the best of our knowledge, this is the first work to explain these anomalies within non-minimal SUSY model, based on extension of MSSM gauge group. In this model, as detailed below, new interacting couplings and new particles are naturally introduced that allow for a new loop contribution with important impacts on $b \to s \ell^+ \ell^- $ anomalies.

The paper is organized as follows. In section 2 we provide the effective Hamiltonian of $b \to s\, \ell^+ \ell^-$ transition and briefly review the general expressions of $R_K$ and $R_{K^*}$ with NP effects. The BLSSMIS model is introduced in section 3, with emphasis on lightest right-handed sneutrino mass and mixing. Section 4 is devoted for new right-handed sneutrino contribution to $R_K$ and $R_{K^*}$. Our numerical analysis is presented in section 5. Finally our conclusions and prospects are give in section 6.   
%%%%%%%%%%%%%%%%%%%%%%%%%%%%%%%
\section{Effective Hamiltonian and $R_K/R_{K^*}$ expressions}
The effective Hamiltonian for $b \to s\, \ell^+ \ell^-$ transition can be written as
\bea
\!\!H_{eff} &=& \sum_i ( C_i (\mu_b) Q_i(\mu_b) \!+\! \tilde{C}_i (\mu_b)\tilde{Q}_i (\mu_b)) + h.c.,~~~
\eea
where $Q_i(\mu_b)$ are the $\Delta B = 1$ transition operators, evaluated at the renomalization scale
$\mu_b \simeq {\cal O}(m_b)$. The relevant operators for our process are given by
\bea
Q_{7\gamma} &=&\frac{m_b}{e} (\bar{s}\sigma^{\mu \nu} P_R b) F_{\mu \nu}, \\
Q_{9} &=& (\bar{s}\gamma^{\mu} P_L b) (\bar{\ell}\gamma^{\mu} \ell), \\
Q_{10} &=&(\bar{s}\gamma^{\mu} P_L b) (\bar{\ell}\gamma^{\mu} \gamma_5 \ell).
\eea
The operators $\tilde{Q}_i$ and Wilson coefficients $\tilde{C}_i $ are obtained from $Q_i$ and $C_i$, respectively, by replacing $L \leftrightarrow R$. In the SM, the electromagnetic dipole operator $Q_{7\gamma}$ and semi-leptonic operators $Q_{9,10}$ give the leading contributions to $b \to s\, \ell^+ \ell^-$. The Wilson coefficients $C_i(\mu)$ at a lower scale $\mu_b = {\cal O}(m_b)$ can be extrapolated by the corresponding ones at high scale $\mu_W ={\cal O}(m_W)$ as

\be
C_i(\mu_b) = \sum_j \hat{U}(\mu_b, \mu_W)_{ij} C_j (\mu_W),
\ee
where the evolution matrix $\hat{U}(\mu_b, \mu_W)$ is given in Ref.\cite{Buras:1994dj,Buchalla:1995vs}.
The numerical values of the SM Wilson coefficients for both $b \to s \mu^+ \mu^-$ and $b\to s e^+ e^-$, corresponding to the central values of SM parameters are given at the  Electro-Weak (EW) scale by \cite{Ali:1999mm}: $C_{7\gamma}=-0.17 \times 10^{-8}$, $C_9 = 0.1 \times 10^{-8}$ and $C_{10} = -0.39\times 10^{-8}$. These coefficients lead to ${\rm BR}(B^+ \to K^+ \mu^+ \mu^-) = {\rm BR}(B^+ \to K^+ e^+ e^-)= 1.1 \times 10^{-7}$, hence $R_K^{\rm SM}=1$.

With NP effects in $b \to s\, \ell^+ \ell^-$, $R_K$ and $R_{K^*}$ can be written as follows \cite{Hiller:2014ula}:
\bea
R_K &\simeq& 1 + \Delta_+,\\
R_{K^*} &\simeq& 1 + \Delta_+ + p ~(\Delta_- - \Delta_+),
\eea
where $\Delta_{\pm}$ are defined by
\bea
\Delta_{\pm} = \frac{2}{\vert C^{\rm SM}_9\vert^2 + \vert C^{\rm SM}_{10}\vert^2} \left[{\rm Re}\left(C^{\rm SM}_9 (C_9^{{\rm NP},\mu} \pm
C_9^{'\mu})^*\right) 
+{\rm Re}\left(C^{\rm SM}_{10} (C_{10}^{{\rm NP},\mu} \pm
C_{10}^{'\mu})^*\right)-(\mu \to e) \right].
\eea
The parameter $p$ is function of $q^2_{\rm min}$ and $q^2_{\rm max}$, such that $p(1\,{\rm GeV}^2, 6\,{\rm Gev}^2) \sim 0.86$ \cite{Hiller:2014ula}. Here we assumed $C^{\rm NP}_i \ll C^{\rm SM}_i$, so only linear terms of $C^{\rm NP}_i/C^{\rm SM}_i$ is kept in the expressions of $R_K$ and $R_{K^*}$.

%%%%%%%%%%%%%%%%%%%%%%%%%%%%%%%
\section{$B-L$ supersymmetric Standard Model with Inverse Seesaw}
The TeV scale BLSSMIS is based on the gauge group $SU(3)_C\times SU(2)_L\times U(1)_Y\times
U(1)_{B-L}$, where the $U(1)_{B-L}$ is spontaneously broken by
chiral singlet superfields $\hat{\eta}_{1,2}$ with $B-L$ charge $=\pm 1$ \cite{Khalil:2015naa}.
A gauge boson $Z'$ and three chiral singlet
superfields $\hat{\nu}_i$ with $B-L$ charge $=-1$ are introduced for
the consistency of the model. Finally, three chiral singlet
superfields $\hat{s}_1$ with $B-L$ charge $=+2$ and three chiral singlet
superfields $\hat{s}_2$ with $B-L$ charge $=-2$ are considered to
implement the inverse seesaw mechanism \cite{Khalil:2010iu}. The particle content of this model, as well as the different charge assignments of each superfield,  is provided in Table \ref{particle-content-BLSSMIS}. 
\begin{center}
\begin{table}[!h]
\begin{tabular}{||c||c|c|c|c|c|c|c|c|c|c|c|c||}\hline\hline
   & $\hat{Q}_i$ & $\hat{U}_i^c$ & $\hat{D}_i^c$ & $\hat{L}_i$ & $\hat{E}_i^c$ & $\hat{\nu}_i^c$ & $\hat{s}_1$ & $\hat{s}_2$ & $\hat{H}_1$ & $\hat{H}_2$ & $\hat{\eta}_1$ & $\hat{\eta}_2$\\
\hline\hline
$SU(3)_c$ & $3$ & $\bar{3}$ & $\bar{3}$ & $1$ & $1$ & $1$ & $1$ & $1$ & $1$ & $1$ & $1$ & $1$\\
\hline
$SU(2)_L$ & $2$ & $1$ & $1$ & $2$ & $1$ & $1$ & $1$ & $1$ & $2$ & $2$ & $1$ & $1$\\
\hline
$U(1)_Y$ & $\frac{1}{6}$ & $-\frac{2}{3}$ & $\frac{1}{3}$ & $-\frac{1}{2}$ & $1$ & $0$ & $0$ & $0$ & $-\frac{1}{2}$ & $\frac{1}{2}$ & $0$ & $0$ \\
\hline
$U(1)_{_{B-L}}$ & $\frac{1}{3}$ & -$\frac{1}{3}$ & -$\frac{1}{3}$ & $-1$ & $1$ & $1$ & $2$ & $-2$ & $0$ & $0$ & $1$ & $-1$\\
\hline\hline
\end{tabular}
\caption{Chiral superfields of the BLSSMIS and their quantum numbers under $SU(3)_C \times SU(2)_L \times U(1)_Y \times U(1)_{B-L}$.}
\label{particle-content-BLSSMIS}
\end{table}
\end{center}
The superpotential of this model is given by%
\bea
W =  Y_u\hat{Q}\hat{H}_2\hat{U}^c + Y_d \hat{Q}\hat{H}_1\hat{D}^c+ Y_e\hat{L}\hat{H}_1\hat{E}^c+Y_\nu\hat{L}\hat{H}_2\hat{\nu}^c
+ Y_s\hat{\nu}^c\hat{\eta}_1\hat{s}_2 +\mu\hat{H}_1\hat{H}_2+ \mu'\hat{\eta}_1\hat{\eta}_2.
\label{superpotential}
\eea
Here, as in the MSSM, we assume that $W$ is invariant under $R$-parity, which is defined as $(-1)^{3(B-L) +2 s}$, so that the lightest SUSY particle is stable. We also assume that the superfields  $\hat{\nu}$, $\chi_{1,2}$  and $\hat{s}_2$ are even under matter
parity, while $\hat{s}_1$ is an odd particle, so that a large mass term $M \hat{s}_1 \hat{s}_2$ is prevented~\cite{Khalil:2015naa}.
The soft SUSY breaking terms are given by
\begin{eqnarray}
- {\cal L}_{\rm soft} &=& m^2_{\tilde{q}ij} \tilde{q}^*_i \tilde{q}_j + m^2_{\tilde{u}ij} \tilde{u}^*_i \tilde{u}_j + m^2_{\tilde{d}ij} \tilde{d}^*_i \tilde{d}_j+ m^2_{\tilde{l}ij} \tilde{l}^*_i \tilde{l}_j
+ m^2_{\tilde{e}ij} \tilde{e}^*_i \tilde{e}_j + m^2_{H_2} \vert H_2 \vert^2 + m^2_{H_1} \vert H_1 \vert^2 +m^2_{\eta_1} \vert{\eta_1}\vert^2 \nonumber\\ 
&+& m^2_{\eta_2}\vert{\eta_2}\vert^2 +m_{\tilde{\nu}}^{2}{\tilde{\nu}}_{i}^{c*}{\tilde{\nu}}_{j}^{c} + m_{\tilde{s_2}}^{2}\tilde{s_2}_{i}^{c*} \tilde{s_2}_{j}^{c} + \bigg{[} Y_{u ij}^{A} \tilde{q}_i \tilde{u}_j H_2 +Y_{d ij}^{A} \tilde{q}_i \tilde{d}_j H_1 + Y_{e ij}^{A} \tilde{l}_i \tilde{e}_j H_1+Y_{\nu ij}^{A}{\tilde{L}}_{i}
{\tilde{\nu}^c}_{j}H_2\nonumber\\
&+& Y_{s ij}^{A}{\tilde{\nu}}_i^{c}{\tilde{s_2}}_j^{c}\eta_{1} +B \mu H_2 H_1+ B \mu^\prime \eta_1 \eta_2 + \frac{1}{2} M_a \lambda^a \lambda^a + M_{BB'} \tilde{B} \tilde{B'} + h.c.
\bigg{]},\nonumber %
\label{Lsoft}%
\end{eqnarray}%
where $(Y_f^A)_{ij} = (Y_f)_{ij} A_{ij}$, the tilde denotes the scalar components of the chiral superfields as well as the fermionic components of the vector superfields and $\lambda^a$ are fermionic components of the vector superfields. For more details of the BLSSMIS and the minimization of the corresponding scalar potential, see Ref.\cite{Khalil:2015naa}. 

The $B-L$ symmetry is radiatively broken by
the non-vanishing VEVs 
 $\langle{\rm Re} \eta^0_i\rangle=\frac{v'_i}{\sqrt{2}}$ ($i=1,2$)
while the EW one by the non-zero
VEVs $\langle{\rm Re} H_{u,d}^0\rangle=v_{u,d}/\sqrt{2}$, with $v=\sqrt{v^2_u+v^d_2}= 246$ GeV, $v'=\sqrt{v'^2_1+v'^2_2}\simeq {\cal O}(1)$ TeV and the ratio of these VEVs are defined as $\tan \beta = v_u/v_d$ and $\tan \beta' = v'_1/v'_2$ \cite{Khalil:2016lgy}. In this case, the conditions for the $(B-L)$ radiative symmetry breaking implies that $\mu'$ can be determined in terms of $M_Z'$ and SUSY soft breaking terms, \ie, $\sim {\cal O}(1)$ TeV.  After $B-L$ and EW symmetry breaking, the neutrino Yukawa
interaction terms lead to the following expression:
\be
{\cal L}_m^{\nu} = m_D\, \bar{\nu}_L \nu^c + M_R\, {\bar {\nu^c}} S_2 +
{\rm {~h.c.}},
\ee
where $m_D=\frac{1}{\sqrt{2}}Y_\nu v_u$ and $ M_R =
\frac{1}{\sqrt 2}Y_{s} v'_1$.
In this framework, the light neutrino masses are related to a
small mass term $\mu_s S^2_2$ in the Lagrangian, with $\mu_s\sim {\cal O}(1)$ {\rm
KeV}, which can be generated at the $B-L$ scale through a
non-renormalisable higher order term $\frac{\chi_1^4 S^2_2}{M^3}$, where $M$  
is the mass of a heavy state whose loop(s) or tree-level tadpole diagrams generate the corresponding higher order term
($M \simeq 10^6$ in our case).Therefore, one finds that the neutrinos mix with the fermionic singlet fields to build up the following $9\times 9$ mass matrix, in the basis
$(\nu_L ,\nu^c, S_2)$:

%
%\textcolor{blue}
\be
{\cal M}_{\nu}= \left(
\begin{array}{ccc}
  0 & m_D & 0\\
  m^T_D & 0 & M_R \\
  0 & M^T_R & \mu_s
\end{array}%
\right). %
\label{inverse}
\ee%
The diagonalisation of the mass matrix, Eq. (\ref{inverse}),
leads to the following light and heavy neutrino masses, respectively: %
\begin{eqnarray}%
m_{\nu_l} &=& m_D M_R^{-1} \mu_s (M_R^T)^{-1} m_D^T,\label{mnul}\\
m_{\nu_h}&=& m_{\nu_{H'}} = \sqrt{M_R^2 + m_D^2}. %
\end{eqnarray} 
Thus, the light neutrino masses can be of order eV, with TeV scale $M_R$, if $\mu_s \ll
M_R$, and order one Yukawa coupling $Y_{\nu}$. Such a large coupling is crucial for testing the BLSSM-IS and probing the heavy (s)neutrinos at the LHC. 
The light neutrino mass matrix
in Eq. (\ref{mnul}) must be diagonalized by the physical neutrino
mixing matrix $U_{\rm MNS}$, {\it i.e.}, %
\be%
U_{MNS}^T m_{\nu_l} U_{MNS} = m_{\nu_l}^{\rm diag} \equiv
{\rm diag}\{m_{\nu_1}, m_{\nu_2}, m_{\nu_3}\}.%
\ee
Therefore, the Dirac neutrino mass
matrix, $m_D$, can be expressed through modified Casas-Ibarra parameterization \cite{Casas:2001sr,Abdallah:2011ew} as follows:
\be
m_D=U_{MNS}\, \sqrt{m_{\nu_l}^{\rm diag}}\, X\, \sqrt{\hat{\mu}_s}^{-1} \hat{M}_R, %
\label{mD}
\ee 
where $X$ is an arbitrary orthogonal matrix. Here, we define the basis for three generations of right-handed sneutrino superfields $\hat{\nu}^c$ and three generations of additional singlet superfields $S_2$ by demanding $M_R$ to be diagonal
$ U^\dagger_S M^T_R U_{\nu^c} = \hat{M}_R^{ii}$  and $\hat{\mu_s}$ is defined as  $\hat{\mu}_s = U^\dagger_S \mu_s U_S$. Note that the rotation matrix $U_S$ is fixed by diagonalizing the mass matrix $\mu_s$. Then one can find $U_{\nu^c}$ that diagonalizes $M_R$ as $U_{\nu^c}^\dagger = U_S^\dagger. M_R^T . (M_R^{\rm diag})^{-1}$, where $M_R^{\rm diag}$ is determined by the eigenvalues of $M_R$.

The
matrix $V$ that diagonalizes the $9\times 9$ neutrino mass matrix
${\cal M}_\nu$, {\it i.e.}, $V^T {\cal M}_\nu V = {\cal
M}_\nu^{\rm diag}$, is given by 
\be V=
\left(%
\begin{array}{cc}
  V_{3\times 3} & V_{3\times6}\\
  V_{6\times 3} & V_{6\times6}  \\
\end{array}%
\right),%
\ee%
with the matrix $V_{3\times3}$ is given by  $V_{3\times3} \simeq \left(1 -\frac{1}{2} F F^T \right)
U_{\rm MNS}\approx U_{\rm MNS}$, where $ F = m_D M^{-1}_R$.  This ensures that the mixing matrix $U_{MNS}$ remains almost unitary, up to small deviation of order $\lsim {\cal O}(10^{-2})$. The matrix $V_{3\times6}$ is defined as $ V_{3\times6}=\left({\bf 0}_{3\times3},F \right) V_{6\times6}$.
Finally, $V_{6\times 6}$ is the matrix that diagonalize the
$\{\nu_R, S_2\}$ mass matrix. The spectrum of heavy neutrinos is determined by the diagonal mass matrix, $M_R$, whose elements are  free  parameters, of order TeV and can be in normal or inverted hierarchy.

We now turn to consider the right-handed sneutrino spectrum, as  the left-handed sneutrino sector almost remains as in the MSSM. If we write $\tilde{\nu}_{R}$ and $\tilde{S}_2$ (the scalar components of the superfields $\hat{\nu}$ and $\hat{s}_2$) as $\tilde{\nu}_R=\frac{1}{\sqrt{2}} ( \tilde{\nu}_R^+ + i\; \tilde{\nu}_R^- )$ and 
$\tilde{S}_2 =  \frac{1}{\sqrt{2}} ( \tilde{S}_2^+ + i\; \tilde{S}_2^-)$, then the CP-even/odd sneutrino mass matrix is given by \cite{Khalil:2011tb,Khalil:2015naa}
%{\small\fontsize{8}{8}\selectfont{ 
\be
{\cal M}^2_{\pm}=\mat{
m_{\tilde{\nu}_R}^2+m_D^2+M_R^2-\frac{1}{2}M_{Z'}^2\cos 2\beta' & \pm M_R (A_S+\mu'\cot\beta')\\\\
\pm M_R (A_S+\mu'\cot\beta') & m_{\tilde{S}_2}^2+M_R^2+M_{Z'}^2\cos 2\beta'
},
\label{Mpm}
\ee
%}}
where $m_{\tilde{\nu}_R}^2$ and $m_{\tilde{S}_2}^2$ are the soft scalar mass matrices and $A_{S}$ is the trilinear coupling, which is also a ($3\times 3$)-matrix. In our analysis, for simplicity, we assume that these matrices are diagonal. The mass eigenvalues of ${\cal M}^2_{\pm}$ are given by 
\begin{eqnarray}
m^2_{\tilde{\nu}_{\mp}} &=& \frac{1}{2}(m_{\tilde{\nu}_R}^2+m_{\tilde{S}_2}^2+m_D^2)+M_R^2+\frac{1}{4}M_{Z'}^2\cos 2\beta' \nonumber\\
&\mp& \sqrt{(m_{\tilde{\nu}_R}^2-m_{\tilde{S}_2}^2+m_D^2)+M_R^2-\frac{3}{2}M_{Z'}^2\cos 2\beta')^2 + 4M_R (A_S+\mu'\cot\beta')}.
\label{eq:mass_splitting}
\end{eqnarray}
Therefore, if $\mu'$ and $A_S$ are of order $m_{\tilde{\nu}_R}$ and $M_R$, \ie, $\sim {\cal O}(1)$~TeV, then the eigenvalue $m^2_{\tilde{\nu}_{-}}$ could the lightest sneutrinos, of order ${\cal O}(100)$~GeV \cite{Abdallah:2017gde}. It can be even the lightest supersymmetric particle and becomes a viable candidate of dark matter \cite{Abdallah:2017gde}.  It is a feature of BLSSM  with Inverse seesaw that stems from the large values of $Y_\nu$. Note that due to suppressed couplings of right-handed sneutrino with SM particles, it survives the current LHC constraints, which are essentially imposed on the left-handed sneutrinos. We call the lightest sneutrino as $\tilde{\nu}_1$ although it is the fourth sneutrino in the mass eigenstate list. 
Finally, the decomposition of $\tilde{\nu}_1$ is given by  
\be 
\tilde{\nu}_1 = \sum_{j=1}^3 Z^R_{1j} (\tilde{\nu}_R)_j + \sum_{k=1}^3 Z^R_{1k} (\tilde{S}_2)_k , ~~~ ~~
\ee
where $ Z^R_{1j}=\frac{1}{\sqrt{2}}(1,0,0)$ and  $ Z^R_{1k}=\frac{1}{\sqrt{2}}(1,0,0)$.\\

%%%%%%%%%%%%%%%%%%%%%%%%%%%%%%%
\section{New BLSSMIS contributions to $R_K/R_{K^*}$}
In the BLSSMIS, the SUSY contribution to $b \to s \ell^+ \ell^-$ can be dominated by the box diagram shown in Fig. \ref{fig:box}, where charginos, light stop, and the lightest right-handed sneutrino are exchanged. Other heavy right-handed sneutrinos, included in our numerical analysis,  will have a negligible effect.  SUSY contributions through penguin diagrams, have subdominant effects and do not lead to any lepton non-universality. One can show that this box diagram may generate scalar, vector, and tensor effective operators. However with neglecting Yukawa couplings of light fermions (keeping only top and neutrino Yukawa couplings), one finds that the leading contributions to the Wilson coefficients $C_9$ and $C_{10}$ are given by:
\begin{figure}[ht]
\centering
\includegraphics[scale=1.5]{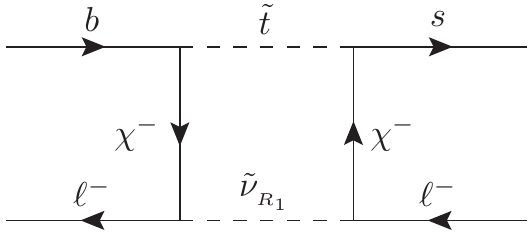}
\caption{The box diagram contributing $b \to s \ell^+ \ell^-$ decay in the BLSSMIS, with charginos, light stop and the lightest right-handed sneutrino exchange. 
\label{fig:box}}
\end{figure}
\bea
\!\!C_9 \!&=&\! (B_{LL} \!+\! B_{LR}),~~~ C'_9 \!=\! (B_{RR} \!+\! B_{RL}),\\
C_{10} \!&=&\!(B_{LL} \!-\! B_{LR}), ~~~ C'_{10} \!=\! (B_{RR} \!-\! B_{RL}),~~
\eea
where $B_{RR}=B_{LR}=B_{RL}=0$, while $B_{LL}$ is given by
\bea
 B^\ell_{LL}= -\Gamma^{d_i \tilde{\chi}_j^+ \tilde{u}_k^*}_L ~ \Gamma^{\bar{d}_i \tilde{\chi}_j^- \tilde{u}_k}_R ~ \Gamma^{\bar{\ell}_i \tilde{\chi}_j^- \tilde{\nu}_k^R}_R ~ \Gamma^{\ell_i \tilde{\chi}_j^+ \tilde{\nu}_k^R}_L 
\times  D_{27}(m^2_{\tilde{\chi}_i^-},m^2_{\tilde{\chi}_j^-},m^2_{\tilde{u}_k}, m^2_{\tilde{\nu}^R_l}).
\eea
The non-vanishing of $B_{LL}$, only, can be understood from the fact that our box contribution, mediated essentially by $\tilde{\nu}_R$ and $\tilde{t}_R$ must have left-handed external fermions, so that it becomes proportional to the large Yukawa couplings $Y_{\nu}$ and $Y_t$. In this respect, the charginos involved in this box diagram should be higgsino-like.    

The non-vanishing couplings (in our approximation) are given by
\begin{eqnarray}
\Gamma^{\bar{\ell}_i \tilde{\chi}_{j}^{-}\tilde{\nu}^{R}_{k}}_{R} = -\frac{1}{\sqrt{2}}\Big[g_2 \sum_{a=1}^3 Z^{R}_{ka} U^\ell_{L,ia} V_{j1}
- \sum_{b=1}^3 \sum_{a=1}^3 Y^*_{\nu,ab} Z^{R}_{k3+a} U^{\ell}_{L,ib} V_{j2}\Big], \label{Gammal}\\
\Gamma^{\bar{d}_i \tilde{\chi}_{j}^{-}\tilde{u}_{k}}_{R} = - \Big[g_2 \sum_{a=1}^3 Z^{u*}_{ka} U^d_{L,ia} V_{j1}
- \sum_{b=1}^3 \sum_{a=1}^3 Y^*_{u,ab} Z^{u*}_{k3+a} U^d_{L, ib} V_{j2}\Big], \label{Gammaq}
\end{eqnarray}
with $\Gamma_{L}=\Gamma^*_R$. Here we assume that quark mixing matrix is given by $U^d_L= V_{CKM}$ and the lepton mixing matrix is given by  $U^\ell_L=U_{MNS}$. Also we assume that the right-handed sneutrino and up-squark mass matrices are diagonalized by $Z^R$ and $Z^u$, respectively, and chargino mass matrix is diagonalized by $U$ and $V$. The loop function $D_{27}(x_i,x_j,x_k,x_l)$ is given by \cite{Abada:2014kba}
\bea
D_{27}(x_i, x_j, x_k, x_l) &=& -\frac{1}{4} \Big[ \frac{x^2_j \log(x_j/x_i)}{(x_j-x_i)(x_j-x_k)(x_j-x_l)} + \frac{x^2_k \log(x_k/x_i)}{(x_k-x_i)(x_k-x_j)(x_k-x_l)} \nonumber\\
&+& \frac{x^2_l \log(x_l/x_i)}{(x_l-x_i)(x_l-x_j)(x_l-x_k)}\Big]\!.~~ ~~
\eea
In this case, we have
\bea
C_9^{\rm SUSY} &=& C_{10}^{\rm SUSY}=B_{LL},\\
C_9^{'{\rm SUSY}}&=&C_{10}^{'{\rm SUSY}}=0.
\eea
Therefore, $\Delta_+$ is given by
\bea
\Delta_+ &\simeq& - 3.578 \times 10^8 ~ (B^\mu_{LL} - B^e_{LL})\nonumber\\
&=&- 3.578 \times 10^8~ \Gamma^{d_i \tilde{\chi}_j^+ \tilde{u}_k^*}_L ~ \Gamma^{\bar{d}_i \tilde{\chi}_j^- \tilde{u}_k}_R
\times\left[\Gamma^{\bar{\mu}_i \tilde{\chi}_j^- \tilde{\nu}_k^R}_R ~ \Gamma^{\mu_i \tilde{\chi}_j^+ \tilde{\nu}_k^R}_L - \Gamma^{\bar{e}_i \tilde{\chi}_j^- \tilde{\nu}_k^R}_R  \Gamma^{e_i \tilde{\chi}_j^+ \tilde{\nu}_k^R}_L \right]\nonumber\\
&\times& D_{27}(m^2_{\tilde{\chi}_i^-},m^2_{\tilde{\chi}_j^-},m^2_{\tilde{u}_k}, m^2_{\tilde{\nu}^R_l}).
\eea

In our numerical analysis, we consider the following ansatz of Yukawa neutrino couplings, which is generated from the expression in Eq.\ref{mD} and is consistent with neutrino experimental data an Lepton Flavor Violation (LFV) constraints:
\be
Y_\nu=   \left(
\begin{array}{ccc}
-0.1179 ~ & ~ 0.01255 ~ &~ 0.00565\\
-0.00765 ~ & ~ 0.024137 ~ & ~ 0.019655 \\
0.002786 ~ & ~ -1.07586 ~ & ~ 0.46574
\end{array}
\right).
\label{Ynu}
\ee
It is noticeable that in this texture the coupling corresponds for first two generations are suppressed ($\lsim 10^{-2}$), so the associated stringent LFV constraints are satisfied. While the entries of third row, which correspond to less constrained LFV of tau decay, could be larger. As we will see, these large Yukawa couplings play important role in enhancing lepton non-universality and reduced $R_K$ to the desired value.  With a proper choice of SUSY parameters at low scale ({\it e.g.}, $M_1=M_2= 1$ TeV, $M_3=2$ TeV, $\tan \beta =45$, $g_{B-L}=5.12$, $\mu \sim \mu'=1.5$ TeV and $v'=4.8$ TeV),  one  gets the following masses and mixing: 

\begin{center}
\begin{table}[!h]
\begin{tabular}{|c|c|c|c|c|c|c|c|c|c|}
\hline\hline
 $m_{\tilde{\nu}^R_1}$  &  $m_{\tilde{\chi}^-_1} $  &  $m_{\tilde{\chi}^-_2} $  &  $m_{\tilde{t}_1}$  &   $Z^R_{13}$   &  $Z^R_{16}$ & $Z^u_{33}$ & $Z^u_{36}$ & $V_{12}$  & $V_{22}$\\
 \hline
 $427.5$ GeV& $1$ TeV & $1.5$ TeV & $458$ GeV & $-0.44$ & $0.89$ & $-0.325$ & $-0.95$ & $- 0.1$ &~ $0.99$ \\
\hline\hline
\end{tabular}
\caption{Benchmark point for $R_K \simeq 0.77$}
\label{example}
\end{table}
\end{center}
From this example, it clear that higgsino-like chargino is favored for enhancing the lepton non-universalit.   
In particular, one finds 
\bea 
\Gamma^{\bar{b} \tilde{\chi}_1^- \tilde{t}_1}_R &\simeq &  Y_t Z^u_{36} (V_{CKM})_{33} V_{12} \sim  0.09, \\
\Gamma^{\bar{b} \tilde{\chi}_2^- \tilde{t}_1}_R &\simeq&  Y_t Z^u_{36} (V_{CKM})_{33} V_{22} \sim -0.9,
\eea
and
\be 
\Gamma^{\bar{s} \tilde{\chi}_2^- \tilde{t}_1}_R &\simeq&  Y_t Z^u_{36} (V_{CKM})_{23} V_{22} \sim 0.01.
\ee
In addition,  $\Gamma^{\bar{\mu}(\bar{e})\tilde{\chi}_2^- \tilde{\nu}_1}_R$  is given by
\bea 
\Gamma^{\bar{\mu}(\bar{e})\tilde{\chi}_2^- \tilde{\nu}_1}_R &\simeq & \frac{1}{\sqrt{2}}  \sum_{b=1}^3 (Y_\nu)_{3b} Z^R_{16} (U_{MNS})_{2(1)b} V_{22} \sim  -0.64(-0.14).
\eea
Recall that the right-handed neutrino has a very small coupling with $SU(2)_L$ gauge interactions, therefore the first two terms in Eq. (\ref{Gammal},\ref{Gammaq}) give negligible contributions.
 Therefore,
\be
\Gamma^{\bar{\mu} \tilde{\chi}_1^- \tilde{\nu}_1^R}_R ~ \Gamma^{\mu \tilde{\chi}_1^+ \tilde{\nu}_1^R}_L - \Gamma^{\bar{e} \tilde{\chi}_1^- \tilde{\nu}_1^R}_R  \Gamma^{e \tilde{\chi}_1^+ \tilde{\nu}_1^R}_L = 0.389. ~~~~~
\ee
Finally for the above mentioned spectrum, one finds that $D_{27}(m^2_{\tilde{\chi}_2^-},m^2_{\tilde{\chi}_2^-},m^2_{\tilde{t}_1}, m^2_{\tilde{\nu}^R_1}) \simeq - 1.6 \times 10^{-7}$. Thus $\Delta_+$ is given by
\bea
\Delta_+ &=& (-3.58 \times 10^8) (-0.9)(0.01)(0.367)(-1.6 \times 10^{-7}) \nonumber\\
&\simeq& - 0.23.
\eea
This leads to
\be
R_K \simeq 0.77,
\ee
which is consistent with the recent measurement reported in Eq.(\ref{RKexp}). It is worth noting that as $C'_9=C'_{10}=0$, we have $\Delta_+=\Delta_-$ and hence in our scenario $R_{K^*} = R_{K} = 1+ \Delta_+$. So we have also $R_{K^*}=0.78$, which lies within $1\sigma$ of the LHCb results given in Eq.(\ref{RK*exp}).

As shown below, we can also account for $R_K$ with a heavier SUSY spectrum, however, it is worth mentioning that the light masses considered in the above example are still allowed by the latest LHC limits~\cite{Belanger:2015vwa}. In particular, light stop $\sim 250$ GeV is allowed if the lightest neutralino is rather heavy, so that the decay $\tilde{t}_1 \to t + \tilde{\chi}_1^0$ is off-shell. In addition, if the chargino mass is degenerate with lightest neutralino mass, then the  decay $ \tilde{\chi}^\pm_1 \to W^\pm \tilde{\chi}_1^0$ is off-shell 
and the decay of stop into chargino via the process 
$ \tilde{t} \to b \tilde{\chi}^\pm_1 \to b W^\pm \tilde{\chi}_1^0$ is suppressed and the experimental exclusion limits could be evaded. 

It worth nothing that the new BLSSM-IS contributions to $R_K$ and $R_{K^*}$ discussed above are  based on the box 
diagram mediated by light right-handed sneutrino and higgsino like chargino with large neutrino Yukawa coupling. As can be seen from Eq.(\ref{Gammal}), the interaction between muon (electron)-higgsino-lightest right handed sneutrino is proprtional to $\sum_b (Y_\nu)_{3b} (U_{MNS})_{2(1)b}$, as the largest mixing of right-handed sneutrino, $Z^R_{1,3+a}$, corresponds to $a=1~ {\rm or}~3$, while the lepton mixing is of the same order for all $b=1,2,3$. Therefore, the difference between muon and electron results in $BR(b \to s \ell^+ \ell^-)$ is mainly due the difference between $$ \sum_b (Y_\nu)_{3b} (U_{MNS})_{2b}~~~ {\rm and}~~~~ \sum_b (Y_\nu)_{3b} (U_{MNS})_{1b}.$$  
Since $ (Y_\nu)_{32} (U_{MNS})_{22}\simeq (Y_\nu)_{32} (U_{MNS})_{12}$, the main difference is coming from 
$$  (Y_\nu)_{33} (U_{MNS})_{23} ~~~~ {\rm vs.} ~~~~~  (Y_\nu)_{33} (U_{MNS})_{13}.$$ Since $(U_{MNS})_{13} < (U_{MNS})_{23}$, the box contribution to $b \to s e^+ e^-$ is smaller than the contribution to $b \to s \mu^+ \mu^-$, hence the observed lepton non-universality can be accommodated. These ingredients would 
require extending the MSSM with two right-handed superfields, so that one can implement the  inverse seesaw mechanism with large $Y_\nu$ and light right-handed sneutrino. The B-L SUSY model, BLSSM-IS, is an interesting example of this extensions, motivated by enlarge the SM gauge group with an extra $U(1)_{B-L}$ symmetry.

%%%%%%%%%%%%%%%%%%%%%%%%%%%%%%%
\section{Numerical Results}
\begin{figure}[h]
\centering
\includegraphics[width=7.5cm,height=5.5cm]{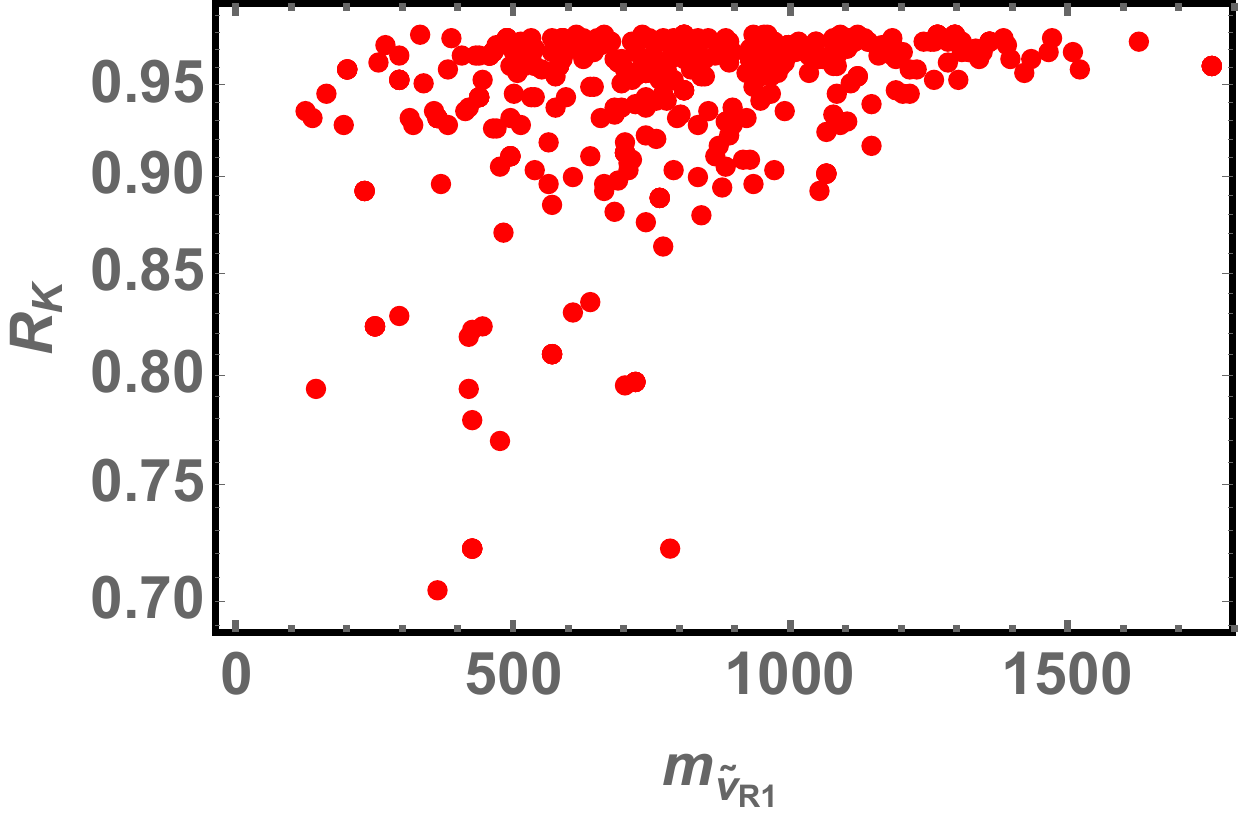}~~~~~\includegraphics[width=7.5cm,height=5.5cm]{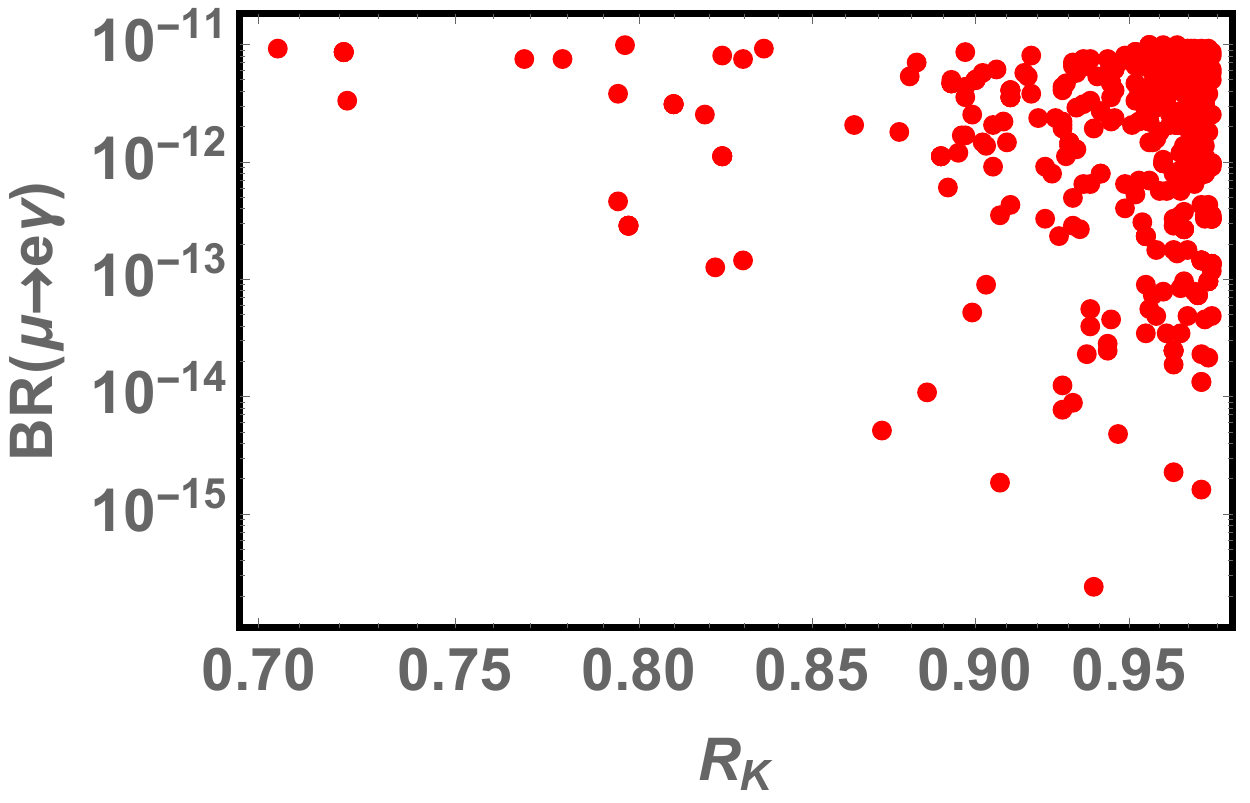}
\caption{(lLeft) $R_K$ versus the lightest right-handed sneutrino mass in the BLSSMIS. (Right) $BR(\mu \to e \gamma)$ versus $R_K$.}
\label{fig_RK_2}
\end{figure}

To confirm our analytical result, we perform comprehensive numerical analysis, based on  FlavorKit \cite{Porod:2014xia}, SARAH \cite{Staub:2013tta} and SPheno
\cite{Porod:2011nf}. We have performed randam scan over the following parameter space
\begin{center}
\begin{table}[!h]
\begin{tabular}{|c|c|}
\hline\hline
 Parmeters  &  Scanned Range\\
 \hline
 $M_1$, $M_2$  & [100,1000] GeV\\
 \hline
 $M_3$  & [2,3] TeV\\
  \hline
 $\tan \beta$  & [10,60]\\
 \hline
Diagonal sfermion masses  & [100, 2000] GeV\\
 \hline
 $A_t$, $A_b$, $A_\tau$, $A_\nu$ $A_s$ & [-2,-1] TeV\\
 \hline
 Other trilinear couplings &0 \\
  \hline
 $M_{H_1}$  & [100,2000] GeV\\
  \hline
 $M_{H_2}$  & [-2000,-100] GeV\\
\hline\hline
\end{tabular}
\caption{Ranges of performed scan}
\label{example}
\end{table}
\end{center}
Moreover, it is assumed that $g_{BL}=0.5$, $\mu_{ii} \simeq 10^{-9}$ GeV, $m_{\eta_1}\sim m_{\eta_2} \simeq 1$ TeV.  
In Fig. \ref{fig_RK_2} (left panel), we display the results of $R_K$ as function of $m_{\tilde{\nu}_R}$ for different values of low energy SUSY parameters, where most of the SUSY particles are quite heavy, except one of the right-handed sneutrino, light stop, and the lightest chargino. Here, we assume $\mu \ll M_2$, so that the lightest chargino is higgsino-like. As can be seen from this figure, 
the right-handed sneutrino  box diagram may give a significant contribution to $R_K$ if $m_{\tilde{\nu}_{R_1}} \lsim 1.5 $ TeV and  also $m_{\tilde{t}_1} , m_{\tilde{\chi}^\pm} \lsim 1$ TeV. It is interesting to note that reducing $R_K$ down to $\sim 0.7$ does not require very light SUSY spectrum. This can be understood from the fact that the suppression in the loop function with heavy masses can be compensated by possible enhancements of the interaction couplings.  It is also notable that, as expected, larger values of $R_K$, \ie, between $0.9$ and 1, are much more plausible. Indeed, $R_K \simeq 0.7$ can be realized at a specific region of the parameter space. However, it is important to note that there is no similar region in the MSSM or other SUSY models. Thus, it is a striking features of BLSSMIS model, with TeV scale right-handed (s)neutrinos and inverse seesaw mechanism. As advocated in the introduction, the SM contribution to $b \to s \ell^+ \ell^-$ is given by one loop penguin diagrams, therefore, it is quite plausible for our right-handed sneutrino box diagram with large couplings to compete the SM effect and gives non-universal contribution to $b \to s \mu^+ \mu^-$ and $b \to s e^+ e^-$ transitions. 

As pointed out in Ref.~\cite{Abdallah:2011ew,Abada:2014kba} that inverse seesaw mechanism with large neutrino Yukawa couplings is stringently constrained by experimental limits of charged LFV, in particular $BR(\mu \to e \gamma) \lsim 10^{-13}$. Therefore, in Fig. \ref{fig_RK_2} (right panel), we plot a correlation between $BR(\mu \to e \gamma) $ and $R_K$, for the same set of input parameters considered in the left panel plot. As can be seen from this figure, the constraints from $\mu \to e \gamma$ limit our results for $R_K$ to be $\gsim 0.8$. However, smaller values of $R_K$ with $BR(\mu \to e \gamma) < 10^{-13}$ is quite plausible, but would require scanning over a wider range of the parameter space and also considering different textures of Yukakw neutrino.

\begin{figure}[!h]
\centering
\includegraphics[width=5.65cm,height=4.5cm]{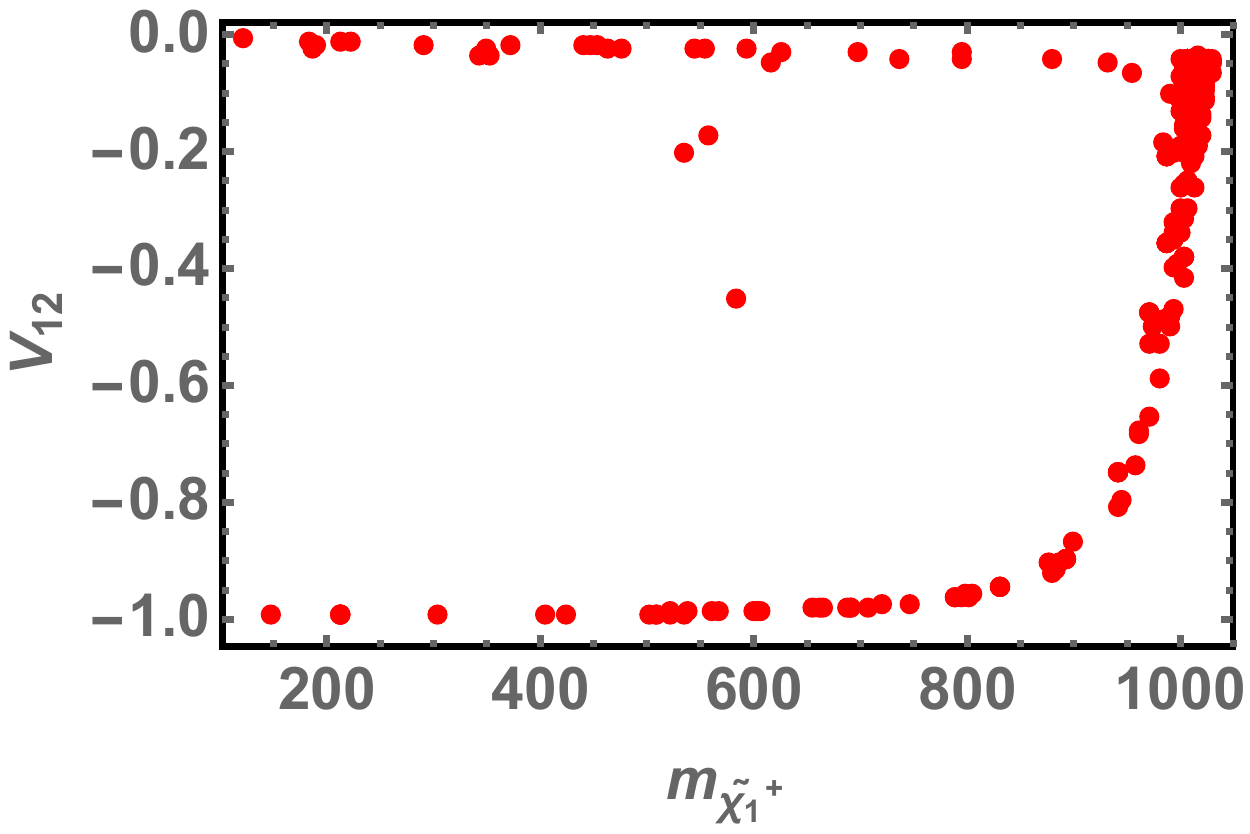}~~~\includegraphics[width=5.65cm,height=4.5cm]{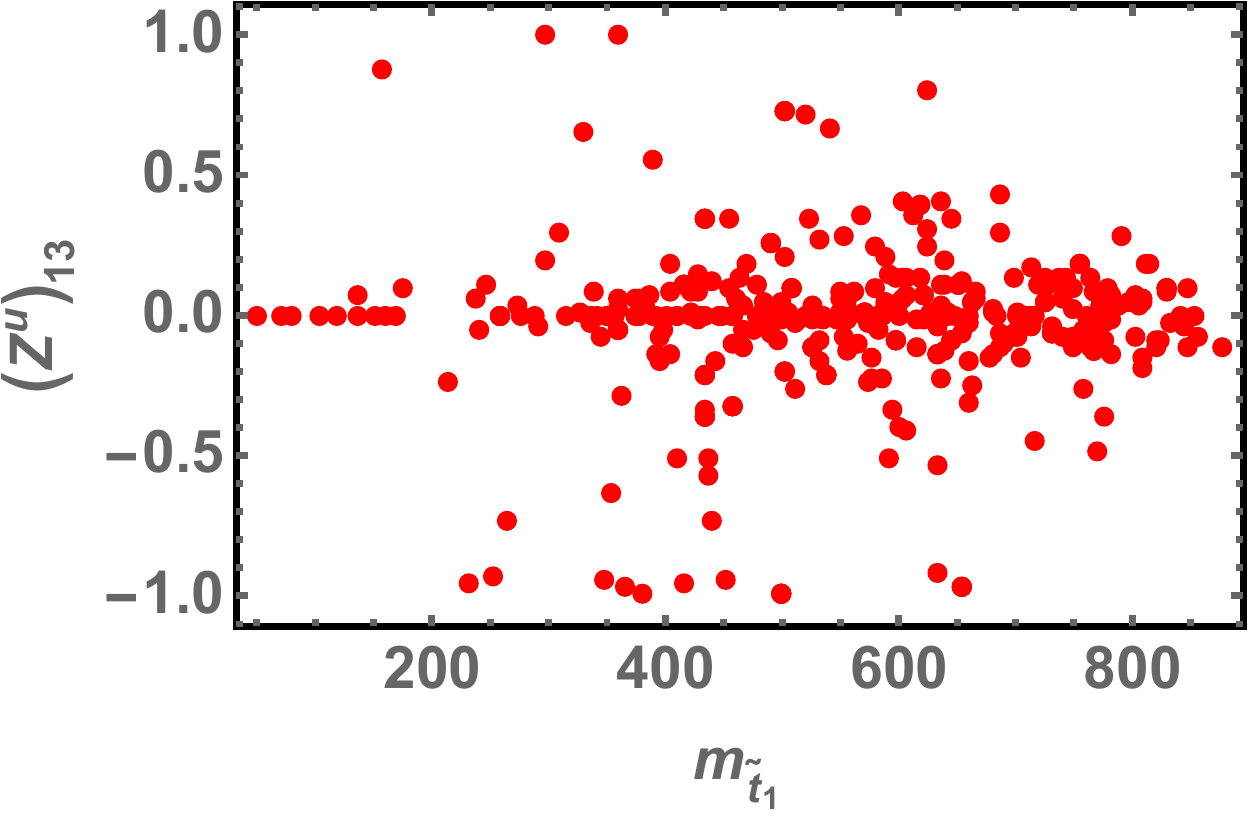}~~~\includegraphics[width=5.65cm,height=4.5cm]{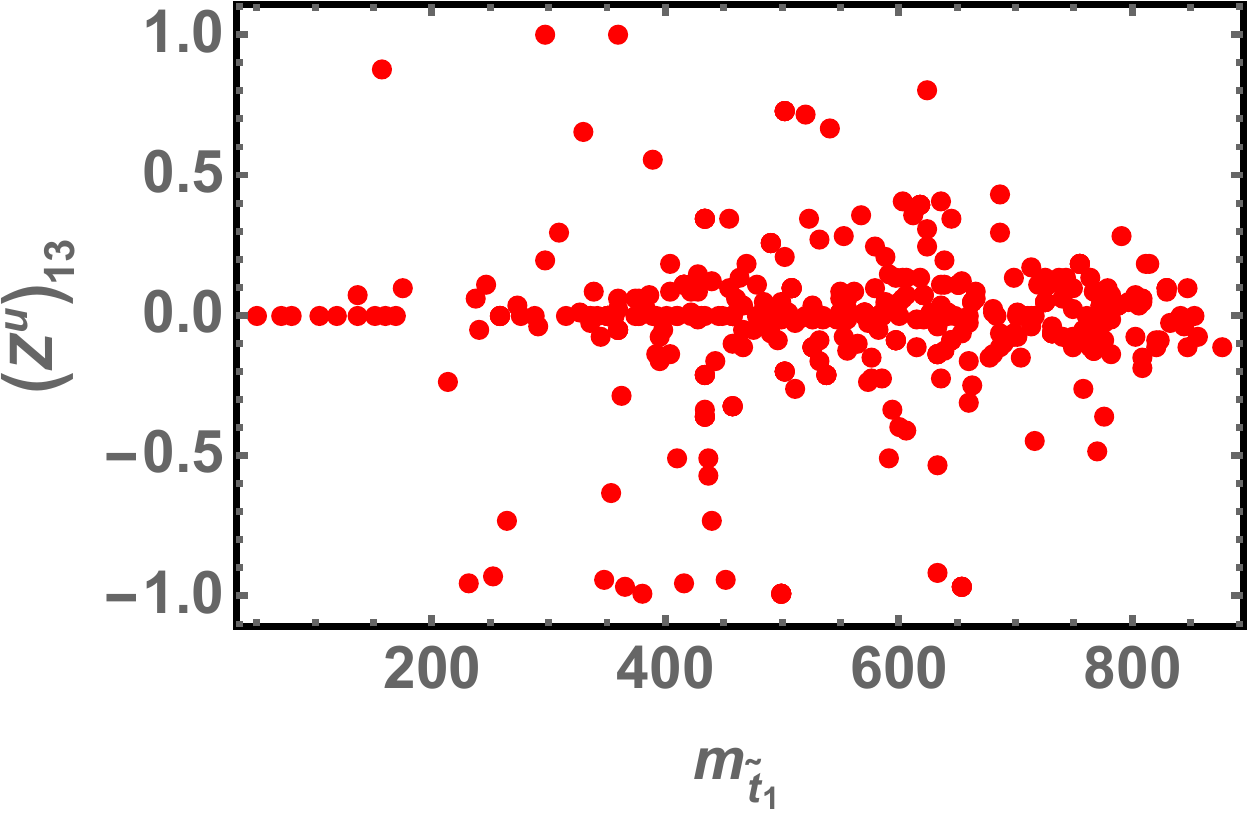}
\caption{(Left) The chargino mixing $V_{12}$ versus the lightest chargino mass. (Middle) The right-handed sneutrino mixing $Z^R_{14}$ versus the lightest sneutrino mass. (Right) The up-squark mixing $Z^u_{13}$ versus the light stop mass.}
\label{fig_mixing}
\end{figure}

In Fig.~\ref{fig_mixing}, we show the relevant mixings, $V_{12}$, $Z^R_{14}$, and $Z^u_{13}$, versus the chargino, lightest sneutrino, and light stop, respectively. One can notice that large mixings can be obtained even if the masses are of order TeV. Those plots are constructed by the same set of results used in Fig.~\ref{fig_RK_2}. From this figure, we aim to show that large mixing  $V_{12}$, $Z^R_{14}$ and $Z^u_{13}$, which are crucial for reducing $R_K$, can be naturally obtained.

In our numerical example, we used modified Casas-Ibarra formula and generated the neutrino Yukawa couplings, $Y_\nu$, as given in Eq. (\ref{Ynu}). This is just an example of set of neutrino Yukawa couplings that can be generated by different values of orthogonal matrix $X$, or the neutrino mas parameters $M_R$ and $\mu_s$. In this example $(Y_\nu)_{23}$ is of order one and this may give an impression that such large coupling is essential for explaining the $R_K$ puzzle, however, as mentioned above, the main reason of such lepton non-universality is the difference between lepton mixing $(U_{MNS})_{23}$ and $(U_{MNS})_{13}$ with large $(Y_{\nu})_{33}$. It is worth stressing that the third generation entries of $Y_\nu$ are not constrained by lepton flavor process, unlike the first two generations. In our analysis, which is based on FlavorKit and Spheno, all constraints from lepton and quark flavor violations, like $b \to s \gamma$, $b \to s \nu \nu$, $B-\bar{B}$ mixing, $\ell_i \to \ell_j \gamma$, $\ell_i \to 3 \ell_j$, etc.,  are implemented and they are naturally satisfied by quite heavy sfermion masses that we consider. As mentioned above, we found that the stringent constraint is due to $\mu \to e \gamma$ and once this constraint is imposed, other constraints automatically satisfied.  

%%%%%%%%%%%%%%%%%%%%%%%%%%%%%%%
\section{Conclusions}
In summary, we have shown that, unlike the MSSM and other SUSY extensions, the BLSSMIS model provides a simultaneous explanation for the recent results of $R_K$ and $R_{K^*}$, which are in clear disagreements with the SM predictions.  We emphasized that the fundamental ingredient in this class of models that allows for the new lepton non-universal results is the existence of a light right-handed sneutrino, with large  non-universal Yukawa couplings with leptons. This particle can generate a new box diagram that provides an important different contributions to $b \to s \ell^+ \ell^-$, with $\ell=\mu$ and $e$. We have shown explicitly, analytically and numerically, that this box diagram has the potential to account for the measured results. Large neutrino Yukawa couplings play a crucial role in 
violating the lepton flavor universality, therefore if these results are confirmed by the forthcoming data, they will be clear signals for new physics not only beyond the SM but also beyond minimal SUSY models. 

Finally, it is worth noting that a Lepton Flavor Violating (LFV) process, like $B \to K \mu^+ e^-$ can also be generated in our model, however, with a lower rate than that of the  lepton conserving process $B \to K \mu^+ \mu^-$, due to the smallness of $(Y_{\nu})_{1i}$ couplings. We find that the branching ratio of this LFV process is about two order of magnitude smaller than its current experimental limit, which is given by ${\rm BR}(B \to K \mu^+ e^-) < 0.7 \times 10^{-6}$. Since the SM prediction for this process is negligibly small, any probe for this process in future experiments would be another smoking gun signature of our model with light RH sneutrino. 

\section*{Acknowledgments}

This work is supported from the STDF project 13858, the European Union's Horizon 2020 research and innovation programme under the Marie Skodowska-Curie grant agreement No 690575, and 
the grant H2020-MSCA-RISE-2014 n. 645722 (Non-Minimal Higgs).
I would like to thank D. Boubaa, S. Salem, and F. Staub for their help.\\

\onecolumngrid

%\twocolumngrid

%%%%%%%%%%%%%%%%%%%%%%%%%%%%%%%%%%%%%%%%%%%%%%%%%%%%%%%%%%%%%%%%%

\end{document}